\documentclass[acmlarge]{acmart}

\setcopyright{none}
\renewcommand\footnotetextcopyrightpermission[1]{}

\usepackage{booktabs} 
\usepackage{hhline}

\usepackage{multicol}
\usepackage{caption}
\usepackage{subcaption}
\usepackage{makecell,shortvrb}
\usepackage{hhline}

\usepackage{xcolor}
\definecolor{P@Blue}{named}{blue}
\definecolor{P@ColorOnBlue}{gray}{.95}
\definecolor{P@GrayFG}{named}{darkgray}
\definecolor{P@GrayBG}{gray}{.90}
\definecolor{P@GrayComment}{gray}{.40}

\RequirePackage{listings}
\lstdefinelanguage{egison}{%
  sensitive = true,
  alsoletter={-,_},
  keywords = [1]{match-all, match-all-dfs, match-first, define, lambda, and, or, not, later, matcher, something, Something, define-macro, let},
  comment=[l]{;},
}%
\lstdefinelanguage{haskell}{%
  sensitive = true,
  comment=[l]{--},
}%
\usepackage{bussproofs}
\usepackage{multicol}
\usepackage[mathletters]{ucs}
\usepackage[utf8x]{inputenc}
\usepackage{here}
\PassOptionsToPackage{hyphens}{url}\usepackage{hyperref}
\usepackage{syntax}
\usepackage{amssymb}
\usepackage{amsmath}
\usepackage{xcolor}
\usepackage{tcolorbox}

\EnableBpAbbreviations

\begin{document}

\setlength{\pdfpageheight}{\paperheight}
\setlength{\pdfpagewidth}{\paperwidth}

\title{Scheme Macros for Non-linear Pattern Matching with Backtracking for Non-free Data Types}

\author{Satoshi Egi}
\affiliation{%
  \institution{Rakuten Institute of Technology, Rakuten, Inc.}
}
\affiliation{%
  \institution{University of Tokyo}
  \country{Japan}
}

\begin{abstract}
  Pattern matching is an important feature of programming languages for data abstraction.
  Many pattern-matching extensions have been proposed and implemented for extending the range of data types to which pattern matching is applicable.
  Among them, the pattern-matching system proposed by Egi and Nishiwaki features practical pattern matching for non-free data types by providing an extensible non-linear pattern-matching facility with backtracking~\cite{egi2018aplas}.
  However, they implemented their proposal only in an interpreter of the Egison programming language, and a method for compiling pattern-matching expressions of Egison was not discussed.
  This paper proposes a method for translating a program that contains pattern-matching expressions of Egison to a program that contains only the existing syntax constructs of functional programming languages.
  This method is based on the three key ideas: (i) transformation of a matcher to a function that takes a pattern and a target, and returns lists of triples of a pattern, a matcher, and a target; (ii) compilation of \texttt{match-all} to application of the \texttt{map} function; (iii) transformation of a value pattern to a function that takes an intermediate pattern-matching result and returns a value.
  This paper shows the proposed method works by showing Scheme macros that provide the users the pattern-matching facility of Egison.
  This paper also presents benchmark results that show Egison pattern-matching embedded in Gauche Scheme is faster than the original Egison interpreter.
\end{abstract}

\ccsdesc[500]{Software and its engineering~General programming languages}


\keywords{macro, pattern matching, non-free data types, non-linear pattern, backtracking}

\maketitle

\section{Introduction}\label{intro}

Pattern matching is an important feature of programming languages for data abstraction.
Pattern matching allows us to replace many verbose function applications for decomposing data (such as \lstinline{car} and \lstinline{cdr}) to an intuitive pattern.
The pattern-matching facility for algebraic data types is common in modern functional programming languages.
For example, Haskell and OCaml support pattern matching for algebraic data types.
Many Scheme implementations also have a similar pattern-matching facility~\cite{gauchePM} though Scheme does not contain pattern matching for algebraic data types in its specification, R7RS~\cite{shinn2013revised}.

However, some data types are not algebraic data types, and we cannot use the common pattern-matching facility for them.
These data types are called \emph{non-free} data types.
Non-free data have no canonical form as data of algebraic data types.
For example, multisets are non-free data types because the multiset $\{a,b,b\}$ has two other equivalent but syntactically different forms $\{b,a,b\}$ and $\{b,b,a\}$.

Many pattern-matching extensions have been proposed for extending the range of data types to which pattern matching is applicable to non-free data types~\cite{Hudak07ahistory,turner2012some}.
Among them, the pattern-matching system proposed by Egi and Nishiwaki~\cite{egi2018aplas} features practical pattern matching for non-free data types with the following three features: (i) non-linear pattern matching with backtracking; (ii) extensibility of pattern-matching algorithms; (iii) polymorphic patterns.
This pattern-matching system is implemented in an interpreter of the Egison programming language.
However, Egi and Nishiwaki did not discuss on a method for efficiently compiling their pattern matching system.
The reason is because it is not as simple as the earlier pattern-matching extensions that can be compiled into simple conditional branches.

This paper proposes a method for translating a program that contains pattern-matching expressions of Egison to a program that contains only the existing syntax constructs of functional programming languages.
The reason why this paper aims to implement the pattern-matching facility of Egison as syntactic sugar of the existing languages is to fully utilize the optimization techniques of an existing compiler without effort for extending an existing compiler.

This paper shows the proposed method works by implementing the pattern-matching facility of Egison on Scheme.
I chose Scheme as the target programming language by the following two reasons:
(i) Scheme has a macro system and provides easy meta-programming environment;
(ii) Scheme is a dynamically typed programming language.
Thanks to (i), we can add a new syntactic construct to Scheme without modifying a Scheme compiler.
I do not use the \lstinline{eval} function for this implementation to fully utilize the optimization techniques of a compiler by handling all the variable bindings natively by a Scheme compiler.
The reason (ii) comes from the problem that the static typing of a program generated by the proposed method is difficult.
The extension of the proposed method for statically typed programming languages remains as future work.
The implementation of the proposed method as Scheme macros has been already open-sourced~\cite{egisonScheme}.
We can try the macros on the Gauche Scheme~\cite{gaucheWeb}.

The remainder of this paper is organized as follows.
Sect.~\ref{history} reviews the history of pattern-matching extensions and implementation of pattern matching in Scheme.
Sect.~\ref{usage} introduces the usage of the proposed Scheme macros for pattern matching.
Sect.~\ref{algorithm} explains the pattern matching algorithm of Egison and implementation of this algorithm in Scheme.
Sect.~\ref{method} explains the implementation of the proposed macros.
Sect.~\ref{performance} discusses the performance of the proposed macros.
Sect.~\ref{future} mentions the remained work.
Finally, Sect.~\ref{conclusion} concludes the paper.

\section{Related Work}\label{history}

This section reviews the history of pattern matching and implementations of pattern matching in Scheme.

\subsection{History of Pattern Matching and Its Extensions}\label{history1}

Pattern matching that looks similar to pattern matching widely used nowadays was proposed by Burstall in 1969~\cite{burstall1969proving}.
Burstall proposed to use the notation ``\lstinline{let cons(a, y) = x}'' instead of ``\lstinline{let (a, y) = decons(x)}''.
In~\cite{burstall1969proving}, \lstinline{concat} is defined in the modern fashion.
Algebraic data types were introduced after pattern matching.
HOPE~\cite{burstall1980hope} proposed in 1980 by Burstall, MacQueen, and Sannella is a well-known language that introduced user-defined algebraic data types and pattern matching for them.

In the 1980s, more expressive pattern matching for a wider range of data types started to be pursued.
Miranda's laws~\cite{thompson1986laws,thompson1990lawful} and Wadler's views~\cite{wadler1987views} are early such research.
They discarded the assumption that one-to-one correspondence should exist between patterns and data constructors.
They enable pattern matching for data types whose data have multiple representation forms.
For example, Wadler's paper~\cite{wadler1987views} presents pattern matching for complex numbers that have two different representation forms: Cartesian and polar.
However, the expressive power of these pattern-matching systems is not enough for representing patterns for non-free data types.
These pattern-matching systems support neither non-linear patterns nor pattern matching with multiple results.

These works in the 1980s lead to the development of more pattern-matching extensions.
Erwig's active patterns~\cite{erwig1996active} proposed in 1996 and Tullsen's first class patterns~\cite{tullsen2000first} proposed in 2000 are such extensions.
Both extensions allow users to customize the pattern-matching algorithm for each pattern constructor.
Active patterns support non-linear patterns though they do not support pattern matching with multiple results.
First class patterns support pattern matching with backtracking though they do not support pattern matching with multiple results.

Egison~\cite{egisonWeb} is a programming language with a pattern matching system that is extensible and supports both of non-linear patterns and multiple results~\cite{egi2018aplas}.
The expressions below match a list ``\texttt{\{1 2 3\}}'' as a list of integers ``\lstinline{(list integer)}'' using the pattern ``\lstinline{<cons $x $ts>}''.
In Egison, a pattern variable begins with ``\lstinline{$}''.
Egison uses four kinds of parenthesis for a different purpose.
``\lstinline{()}'' is used to build a syntax tree or apply a function.
``\lstinline{<>}'' is used to apply a pattern constructor to patterns.
``\lstinline{[]}'' is used to build a tuple.
``\texttt{\{\}}'' is used to build a list. 

\begin{lstlisting}[language=egison]
(match-all {1 2 3} (list integer) [<cons $x $ts> [x ts]]) ; {[1 {2 3}]}
\end{lstlisting}

\noindent
The \lstinline{match-all} expression used in the above program returns a list of the results evaluating the body expression for all the pattern-matching results.
The \lstinline{cons} pattern decomposes a list into the head element and the rest.
Therefore, in this case, the evaluation result of this \lstinline{match-all} contains just a single element.
The pattern-matching expression of Egison takes a \emph{matcher} in addition to a target and match clauses.
A matcher specifies the method to interpret the pattern.
The matcher ``\lstinline{(list integer)}'' specifies that we pattern-match the target with the given pattern as a list of integers.
Users can define their own matchers in Egison.

If we change the matcher of the above \lstinline{match-all} from ``\lstinline{(list integer)}'' to ``\lstinline{(multiset integer)}'', the evaluation result also changes.
The \lstinline{cons} pattern of a multiset is defined to divide a target list into an arbitrary element and the rest.
As a result, there are three decompositions.

\begin{lstlisting}[language=egison]
(match-all {1 2 3} (multiset integer) [<cons $x $ts> [x ts]])
; {[1 {2 3}] [2 {1 3}] [3 {1 2}]}
\end{lstlisting}

In addition to \lstinline{cons}, the \lstinline{join} pattern can be used for the \lstinline{list} matcher.
The \lstinline{join} pattern divides a target list into two lists, a head part and tail part.

\begin{lstlisting}[language=egison]
(match-all {1 2 3} (list integer) [<join $hs $ts> [hs ts]])
; {[{} {1 2 3}] [{1} {2 3}] [{1 2} {3}] [{1 2 3} {}]}
\end{lstlisting}

The expressions below match a list that consists of $n$ zeros as a multiset of integers with patterns that match a sequential triple ``\lstinline{<cons $x <cons ,(+ x 1) <cons ,(+ x 2) _>>>}''.
In the pattern-matching expression below, the target list does not contain a sequential triple.
As a result, this expression returns an empty list.
In Egison, a value pattern that checks equality of data begins with ``\lstinline{,}''.
After ``\lstinline{,}'',  we can write an arbitrary expression.
The time complexity of this pattern matching is $O(n^2)$ because Egison uses backtracking for traversing a search tree.
The Curry functional logic programming language also supports non-linear pattern matching with multiple results.
However, Curry transforms a non-linear pattern to pattern guards~\cite{antoy2010programming,antoy2001constructor,hanus2007multi}.
Therefore, the time complexity for the same pattern matching is $O(n^3)$ in Curry because pattern guards are applied after pattern matching succeeds.
Thus, backtracking is an important feature for handling non-linear patterns.

\begin{lstlisting}[language=egison]
(match-all (take n (repeat 0)) (multiset integer) [<cons $x <cons ,(+ x 1) <cons ,(+ x 2) _>>> x])
; returns [] in O(n^2) time
\end{lstlisting}

\noindent
Of course, if we use a more efficient data structure for lists, we can find a sequential triple more efficiently.
For example, if we assume that the target list is sorted, the time complexity for finding a sequential triple is $O(n)$.
In fact, we can also define such a pattern-matching algorithm by defining a matcher for sorted-lists.

The program below defines a pattern-matching algorithm for a multiset.

\begin{lstlisting}[language=egison]
(define $multiset
  (lambda [$a]
    (matcher
      {[<nil> []
        {[{} {[]}]
         [_ {}]}]
       [<cons $ $> [a (multiset a)]
        {[$tgt (match-all tgt (list a)
                 [<join $hs <cons $x $ts>> [x {@hs @ts}]])]}]
       [,$val []
        {[$tgt (match [val tgt] [(list a) (multiset a)]
                 {[[<nil> <nil>] {[]}]
                  [[<cons $x $xs> <cons ,x ,xs>] {[]}]
                  [[_ _] {}]})]}]
       [$ [something]
        {[$tgt {tgt}]}]})))
\end{lstlisting}

\noindent
The \lstinline{matcher} expression used in the above program is a built-in syntax of Egison for defining pattern-matching algorithms for each data type.
The definition of \lstinline{multiset} in Scheme will be explained in Sect.~\ref{method-matcher}.
I will skip the explanation of the following program for now.
This paper imports the pattern-matching facility of Egison into Scheme.

\medskip

Pattern matching was also invented in the context of computer algebra.
Pattern matching for symbolical mathematical expression was implemented in the symbol manipulation system proposed by McBride~\cite{mcbride1969symbol}, which was developed on top of Lisp.
This pattern-matching system supports non-linear patterns.
Their paper demonstrates some examples that process symbolical mathematical expressions to show the expressive power of non-linear patterns.
However, this approach does not support pattern matching with multiple results, and users cannot extend its pattern-matching facility.

Queinnec~\cite{queinnec1990compilation} also pursued expressive pattern matching.
Though this proposal is specialized in lists and not extensible, the proposed pattern-matching system is as expressive as Egison.
The proposed system supports the \lstinline{cons} and the \lstinline{join} patterns, \lstinline{match-all}, not-patterns, and recursive patterns.


\subsection{Implementation History of Pattern Matching in Scheme}\label{history2}


Currently, a SRFI for pattern matching for algebraic data types does not exist.
However, pattern matching for algebraic data types designed by Wright~\cite{wright1993pattern} is implemented in most of the well-known Scheme implementations such as Gauche~\cite{gauchePM}, Guile~\cite{guilePM}, and Chicken~\cite{chickenPM}.
Among them, Racket~\cite{plt-tr1} provides more expressive pattern matching than other Scheme implementations.
The \lstinline{match} expander~\cite{tobin2011extensible} of Racket allows users arbitrary transformation of data when pattern matching.
For example, we can describe view patterns~\cite{wadler1987views} using the \lstinline{match} expander.

This paper implements more expressive pattern matching that supports non-linear pattern matching with backtracking just by introducing Scheme macros.
This work can be easily ported to other relatives of Scheme.
For example, Yuito Murase has exported the proposed Scheme macros to Common Lisp~\cite{egisonLisp}.

%
%
%
%
%
%
%
%
%

\section{Syntax of the Proposed Macros}\label{usage}

This section introduces the usage of the pattern-matching expressions provided by the proposed Scheme macros.
Fig.~\ref{fig:syntax} shows the formal grammar of the pattern-matching expressions of the proposed Scheme macros.
$M$ denotes an expression for pattern matching.
$e$ denotes an arbitrary expression, including $M$.
$p$ denotes a pattern.
$x$ denotes a symbol.
$C$ denotes a pattern constructor.

\begin{figure}[h]
  \vspace{-3mm}
\begin{multicols}{2}
\noindent
\begin{align*}
M &::= \texttt{(match-all $e$ $e$ [$p$ $e$] $\cdots$)} \\
&\mid \texttt{(match-first $e$ $e$ [$p$ $e$] $\cdots$)} \\
&\mid \texttt{Something}
\end{align*}
\columnbreak
\begin{align*}
p &::= \texttt{_} \mid \texttt{$x$} \mid \texttt{,$e$} \mid \texttt{($C$ $p$ $\cdots$)} \mid \texttt{'($p$ $\cdots$)} \\
&\mid \texttt{(or $p$ $\cdots$)} \mid \texttt{(and $p$ $\cdots$)} \mid \texttt{(not $p$)} \\
&\mid \texttt{(later $p$)} \\
\end{align*}
\end{multicols}
  \vspace{-8mm}
  \caption{Syntax of the proposed macros}
  \label{fig:syntax}
\end{figure}

Scheme programs in this paper use three kinds of parenthesis.
``\lstinline{[]}'' is used to build a tuple.
``\texttt{\{\}}'' is used to build a list. 
``\lstinline{()}'' is used to build a syntax tree or apply a function and a pattern constructor.
Gauche treats these three parentheses in the same way as ``\lstinline{()}''.
Though we use ``\lstinline{<>}'' for applying a pattern constructor in Egison, we use ``\lstinline{()}'' for that in Scheme because we cannot use ``\lstinline{<>}'' as a parenthesis in Scheme.~\footnote{The square brackets ``\lstinline{[]}'' are reserved for future language extension in R7RS.
Furthermore, the use of braces ``\texttt{\{\}}'' is outside the standard language specification in both of R6RS and R7RS.
Therefore, it is safe to replace braces and square brackets to parenthesis when we export the Gauche Scheme program in this paper to other Scheme implementation.
}

The \texttt{match-all} expression has the completely same meaning as that of Egison.
The \texttt{match-first} is similar to the traditional \texttt{match} expression; it evaluates the body of the first match clause whose pattern matches with the target.
We do not use the name \texttt{match} to avoid the name conflict with Wright's \texttt{match}~\cite{wright1993pattern} because Wright's \texttt{match} plays fundamental role for defining a user-defined matcher, which will be explained in Sect.~\ref{method-matcher}.
The only difference between \texttt{match-first} and the traditional \texttt{match} expression is \texttt{match-first} takes a matcher.
The both \texttt{match-all} and \texttt{match-first} expression can take multiple match clauses.
\lstinline{Something} is the only built-in matcher of the Egison pattern-matching system.
\lstinline{Something} can handle only wildcards or pattern variables, and is the only matcher that can bind a value to a pattern variable.

The proposed macros provide several pattern constructs.
The rest of this subsection explains each of them.

\subsection{Wildcard and Pattern Variables}

Symbols that appear in a pattern are handled as pattern variables.
The value assigned to a pattern variable can be referred to in its right side of the pattern.
``\lstinline{_}'' represents a wildcard.
A wildcard is pattern-matched with a target in the same method with a pattern variable though no assignment is created for a wildcard.

\subsection{Value Patterns}

Value patterns are represented by prepending ``\lstinline{,}'' to an expression that is evaluated to a value.
Value patterns are used for expressing non-linear patterns.
The pattern-matching algorithms for value patterns are also defined in matchers.

The program below pattern-matches a list \lstinline{(1 2 5 9 4)} as a multiset.
``\lstinline{,(+ x 1)}'' inside the pattern is a value pattern.
The value assigned to the pattern variable \lstinline{x} appeared on the left side of the pattern can be referred to inside the value pattern.
The pattern in the program below matches if the target collection contains pairs of elements in a sequence.

\begin{lstlisting}[language=egison]
(match-all '(1 2 5 9 4) (Multiset Integer) [(cons x (cons ,(+ x 1) _)) x]) ; (1 4)
\end{lstlisting}

The use of ``\lstinline{,}'' to notate a value pattern is an analogy of the use of ``\lstinline{,}'' in \lstinline{quasiquote}.
A value pattern is similar to \lstinline{unquote} in the sense that only the expressions inside value patterns that are followed after ``\lstinline{,}'' are evaluated as an ordinary Scheme expression in a pattern.

\subsection{Constructor Patterns}

When the pattern is a list, the first element is always handled as a pattern constructor.
For example, \lstinline{join} and \lstinline{cons} that appear in ``\lstinline{(join _ (cons x _))}'' are pattern constructors.
The pattern-matching algorithm for handling them is defined and retained in matchers.
For example, in the \lstinline{List} matcher, \lstinline{join} is defined as a pattern constructor that divides a collection into a head and tail part, and \lstinline{cons} is defined as a pattern constructor that divides a collection into the head element and rest.
As a result, the following pattern matches each element of the target list.

\begin{lstlisting}[language=egison]
(match-all '(1 2 3) (List Integer) [(join _ (cons x _)) x]) ; (1 2 3)
\end{lstlisting}

\subsection{Tuple Patterns}

Tuple patterns are represented by prepending ``\lstinline{'}'' to a list of patterns.
Each element of a tuple pattern is pattern-matched with the corresponding element of a target list using the corresponding element of a matcher list as a matcher.

\begin{lstlisting}[language=egison]
(match-all '[1 2] `[,Integer ,Integer] ['[x y] `(,x ,y)]) ; ((1 2))
(match-all '[1 2 3] `[,Integer ,Integer ,Integer] ['[x y z] `(,x ,y ,z)]) ; ((1 2 3))
\end{lstlisting}

``\lstinline{'}'' is important to distinguish a tuple pattern from a constructor pattern.
For example, ``\lstinline{'[x y]}'' cannot be distinguished from a constructor pattern whose constructor is ``\lstinline{x}'' if ``\lstinline{'}'' is not prepended.

\subsection{Logical Patterns}

Logical pattern constructs, or-patterns, and-patterns, and not-patterns are useful also in Egison.
Its usage is similar to that in the existing languages.
An or-pattern matches if one of the argument patterns matches the target.
An and-pattern matches if all the argument patterns match the target.
A not-pattern matches if the argument pattern does not match the target.

\begin{lstlisting}[language=egison]
(match-all '(1 2 3) (List Integer) [(cons (or ,1 ,10) _) "OK"]) ; ("OK")
(match-all '(1 2 3) (List Integer) [(cons (and ,1 x) _) x]) ; (1)
(match-all '(1 2 3) (List Integer) [(cons x (not (cons ,x _))) x]) ; (1)
\end{lstlisting}

\subsection{Later Patterns}

A later pattern is used to change the order of the pattern-matching process.
Basically, the pattern-matching system of Egison processes patterns from left to right in order.
However, we sometimes want to change this order, for example, to refer to the value bound to the right side of pattern variables.
A later pattern can be used for such purposes.
The pattern inside a later pattern is pattern-matched after pattern matching for the other parts of the pattern has succeeded.

\begin{lstlisting}[language=egison]
(match-all '(1 1 2 3) (List Integer) [(cons (later ,x) (cons x _)) x]) ; (1)
\end{lstlisting}

\section{Algorithm of Egison Pattern Matching and Its Implementation in Scheme}\label{algorithm}

This section explains and implements the pattern-matching algorithm of Egison, which is described in detail in Sect. 5 and Sect. 7 of the original paper of Egison~\cite{egi2018aplas}.
After the explanation of the pattern-matching algorithm, I introduce a simplified implementation of this algorithm in Scheme.
This implementation is called by the proposed macros whose implementation is introduced in Sect.~\ref{method}.

\medskip

First, I start by explaining the pattern-matching algorithm of Egison briefly.
In Egison, pattern matching is implemented as reductions of \emph{matching states}.
A matching state consists of a stack of \emph{matching atoms} and an intermediate result of pattern matching.
A matching atom is a triple of a pattern, target, and matcher.
In each step of a pattern-matching process, the top matching atom of the stack of matching atoms is popped off.
From this matching atom, a list of lists of next matching atoms is calculated.
Each list of the next matching atoms is pushed on the stack of matching atoms of the matching state.
As a result, a single matching state is reduced to multiple stacks of matching states in a single reduction step.
Pattern matching is recursively executed for each matching state.
When a stack becomes empty, it means pattern matching for this matching state succeeded.
Fig.~\ref{fig:reductionPath} shows the reduction path when executing the \lstinline{match-all} expression below.
The matching state that is not grayed-out in the $i$-th row is reduced to matching states in the $(i+1)$-th row.
For example, the first matching state in the second row is reduced to the matching state in the third row.
\texttt{MState} takes two arguments, a stack of matching atoms and an intermediate pattern-matching result.
The value pattern \lstinline{,m} is transformed to \lstinline{(val ,(lambda (m) m))}.
This transformation is explained in Sect.~\ref{method-val-pat}.

\begin{lstlisting}[language=egison]
(match-all '(2 8 2) (Multiset Integer) [(cons m (cons ,m _)) m]) ; (2 2)
\end{lstlisting}

\begin{figure}[t]
\begin{center}
  \footnotesize
  \begin{tabular}{ll}
    \toprule
    1 &  \texttt{(MState \{[(cons m (cons (val ,(lambda (m) m)) \_)) (Multiset Integer) \{2 8 2\}]\} \{\})} \\ \hline
    2 &  \thead[l]{\texttt{(MState \{[m Integer 2] [(cons (val ,(lambda (m) m)) \_) (Multiset Integer) \{8 2\}]\} \{\})} \\
                   \textcolor{gray}{\texttt{(MState \{[m Integer 8] [(cons (val ,(lambda (m) m)) \_) (Multiset Integer) \{2 2\}]\} \{\})}} \\
                   \textcolor{gray}{\texttt{(MState \{[m Integer 2] [(cons (val ,(lambda (m) m)) \_) (Multiset Integer) \{2 8\}]\} \{\})}}} \\ \hline
    3 &  \texttt{(MState \{[m Something 2] [(cons (val ,(lambda (m) m)) \_) (Multiset Integer) \{8 2\}]\} \{\})} \\ \hline
    4 &  \texttt{(MState \{[(cons (val ,(lambda (m) m)) \_) (Multiset Integer) \{8 2\}]\} \{2\})} \\ \hline
    5 &  \thead[l]{\textcolor{gray}{\texttt{(MState \{[,m Integer 8] [\_ (Multiset Integer) \{2\}]\} \{2\})}} \\
                   \texttt{(MState \{[,m Integer 2] [\_ (Multiset Integer) \{8\}]\} \{2\})}} \\ \hline
    6 &  \texttt{(MState \{[\_ (Multiset Integer) \{8\}]\} \{2\})} \\ \hline
    7 &  \texttt{(MState \{[\_ Something \{8\}]\} \{2\})} \\ \hline
    8 &  \texttt{(MState \{\} \{2\})} \\
    \bottomrule
  \end{tabular}
\end{center}
  \caption{Reduction path of matching states}
  \label{fig:reductionPath}
\end{figure}

\medskip

\begin{figure}[t]
 \begin{subfigure}[b]{1.0\linewidth}
\begin{lstlisting}[language=egison]
(define processMState
  (lambda (mState)
    (match mState
      ...
      [('MState {[pvar 'Something t] . mStack} ret)
       `((MState ,mStack ,(append ret `{,t})))]
      [('MState {[p M t] . mStack} ret)
       (let {[next-matomss (M p t)]}
         (map (lambda (next-matoms) `(MState ,(append next-matoms mStack) ,ret)) next-matomss))])))
\end{lstlisting}
  \caption{The \texttt{processMState} function}
  \label{fig:processMState}
 \end{subfigure}
  \medskip
 \begin{subfigure}[b]{1.0\linewidth}
\begin{lstlisting}[language=egison]
(define processMStates
  (lambda (mStates)
    (match mStates
      [() '{}]
      [(('MState '{} ret) . rs)
       (cons ret (processMStates rs))]
      [(mState . rs)
       (processMStates (append (processMState mState) rs))])))
\end{lstlisting}
  \caption{The \texttt{processMStates} function}
  \label{fig:processMStates}
 \end{subfigure}
  \medskip
 \begin{subfigure}[b]{1.0\linewidth}
\begin{lstlisting}[language=egison]
(define processMStates1
  (lambda (mStates)
    (match mStates
      [() '{}]
      [(('MState '{} ret) . rs)
       `(,ret)]
      [(mState . rs)
       (processMStates1 (append (processMState mState) rs))])))
\end{lstlisting}
  \caption{The \texttt{processMStates1} function}
  \label{fig:processMStates1}
 \end{subfigure}
 \caption{Implementation of the Egison pattern matching procedure in Scheme}
\end{figure}

The \lstinline{processMState} function in Fig.~\ref{fig:processMState} implements this process.
\lstinline{processMState} takes a matching state as its argument and returns a list of next matching states.
In the program in Fig.~\ref{fig:processMState}, several matching clauses are omitted for simplification.
``\lstinline{...}'' in the 4th line represents this omission.

The match clause in the 5th and 6th lines describes the case where the matcher of the top matching atom is \lstinline{Something}.
In this case, the value of the target ``\lstinline{t}'' is added to the intermediate pattern-matching result.
The \lstinline{Something} matcher is simply defined as follows.

\begin{lstlisting}[language=egison]
(define Something 'Something)
\end{lstlisting}

The match clause in the 7th-9th lines describes the general case.
The list of lists of next matching atoms is calculated in the 8th line.
In this Scheme implementation, a matcher is a function that takes a pattern and a target, and returns a list of lists of the next matching atoms as explained again in Sect.~\ref{method-matcher}.

In the program in Fig.~\ref{fig:processMState}, the matching clauses for and-patterns, or-patterns, not-patterns, and later patterns introduced in Sect.~\ref{usage} are omitted.
We can implement these pattern constructs by adding match clauses for them.
We can see the full implementation of \lstinline{processMState} in \texttt{egison.scm} in the GitHub repository of the proposed macros~\cite{egisonScheme}.

\medskip

The \lstinline{processMStates} in Fig.~\ref{fig:processMStates} implements the whole pattern-matching process while the above \lstinline{processMState} implements a step of this pattern-matching procedure.
\lstinline{processMStates} takes a list of matching states as its argument and returns all pattern-matching results.
The first match clause (line 4) describes the case where a list of matching states is empty.
In this case, \lstinline{processMStates} returns an empty list and terminates the pattern-matching procedure.
The second match clause (lines 5 and 6) describes the case where the stack of the first matching state is empty.
In this case, the intermediate pattern-matching result of this matching state is added to the return value of \lstinline{processMStates} as a final result of pattern matching.
\lstinline{processMStates} is recursively called for the remaining matching states.
The third match clause (lines 7 and 8) describes the general case.
In this case, \lstinline{processMState} is called for the first matching state.
\lstinline{processMStates} is recursively called for the result of appending the result of this \lstinline{processMState} and the remaining matching states.

The above implementation is from \texttt{egison.scm}~\cite{egisonScheme} that does not support pattern matching with infinitely many results.
In \lstinline{processMStates} of \texttt{stream-egison.scm}~\cite{egisonScheme}, the traversal of an infinitely large search tree is implemented that is explained in Sect.~5.2 of the original Egison paper~\cite{egi2018aplas}.
Appendix~\ref{pm-inf} demonstrates pattern matching that uses the macros provided by \texttt{stream-egison.scm}.

\lstinline{processMStates} strictly evaluates all the pattern-matching results even when only the first pattern-matching result is used.
Consequently, sharing the same \lstinline{processMStates} function with \lstinline{match-first} is inefficient.
For that reason, the \lstinline{processMStates1} function in Fig.~\ref{fig:processMStates1} is used for \lstinline{match-first} instead of \lstinline{processMStates}.
\lstinline{processMStates1} is defined to calculate only the first pattern-matching result.
This distinct implementation of \lstinline{processMStates} is not necessary for the Egison language that adopts a lazy evaluation strategy.


\section{Implementation Method}\label{method}

There are three problems for porting the Egison pattern-matching system into Scheme.
The three tricks explained in Sect.~\ref{method-matcher}, Sect.~\ref{method-map}, and Sect.~\ref{method-val-pat} solve these problems, respectively.

\begin{enumerate}
\item How to represent a matcher in Scheme that is a special built-in object of Egison?
\item How to implement a syntax construct that calls the complicated pattern-matching procedure in Sect.~\ref{algorithm}?
\item How to represent a non-linear pattern that has the complex scoping rule in Scheme?
\end{enumerate}

\subsection{Compiling Matchers to \lstinline{lambda}s}\label{method-matcher}

In the Egison pattern-matching system~\cite{egi2018aplas}, a matcher is a special object that retains a pattern-matching algorithm.
In the proposed Scheme macros, we encode a matcher with \lstinline{lambda}s to avoid introducing new built-in objects.
We can achieve that by defining a matcher as a function that takes a pattern and a target, and returns a list of lists of next matching atoms.
For example, the \lstinline{Multiset} matcher is defined as follows.

\begin{lstlisting}[language=egison]
(define Multiset
  (lambda (M)
    (lambda (p t)
      (match p
        [('nil) (if (eq? t '{}) '{{}} '{})]
        [('cons px py)
         (map (lambda (xy) `{[,px ,M ,(car xy)] [,py ,(Multiset M) ,(cadr xy)]})
              (match-all t (List M)
                [(join hs (cons x ts)) `[,x ,(append hs ts)]]))]
        [('val v)
         (match-first `[,t ,v] `[,(List M) ,(Multiset M)]
           ['[(nil) (nil)] '{{}}]
           ['[(cons x xs) (cons ,x ,xs)] '{{}}]
           ['[_ _] '{}))]
        [pvar `{{[,pvar Something ,t]}}]))))
\end{lstlisting}

\noindent
Note that the separate use of ``\texttt{\{\}}'' and ``\lstinline{[]}'' introduced in Sect.~\ref{usage} enhances the readability of the above program.

As a matcher definition of Egison~\cite{egi2018aplas}, a matcher definition consists of match clauses for pattern matching against a pattern.
Each match clause describes a pattern-matching algorithm against the matched pattern.
For pattern matching against patterns, the traditional Wright's \lstinline{match} expression~\cite{wright1993pattern} is used.

The first match clause (line 5) describes the pattern-matching algorithm for the \lstinline{nil} pattern that matches with an empty list.
When the target list is empty, this match clause returns a list that consists of an empty list of matching atoms.
Otherwise, this match clause returns an empty list.
That means pattern matching fails.

The second match clause (lines 6-9) describes a pattern-matching algorithm for the \lstinline{cons} pattern.
The pattern of this match clause is  ``\lstinline{('cons px py)}''.
The first and second argument of the \lstinline{cons} pattern is assigned to \lstinline{px} and \lstinline{py} respectively.
The body of this match clause is a bit complicated.
The evaluation result of this body expression depends on the target list.
Now, we consider a case where the target list is ``\lstinline{(1 2 3)}''.
In this case, this body expression is evaluated to ``\texttt{\{\{[px M 1] [py (Multiset M) (2 3)]\} \{[px M 2] [py (Multiset M) (1 3)]\} \{[px M 3] [py (Multiset M) (1 2)]\}\}}''.
Each list of the next matching atoms is pushed to the current matching state; as a result, three new matching states are generated.
For the first matching state, \lstinline{1} and \lstinline{(2 3)} are pattern-matched using the ``\lstinline{M}'' and ``\lstinline{(Multiset M)}'' matcher with the patterns \lstinline{px} and \lstinline{py}, respectively, in the succeeding pattern-matching process.

The third match clause (lines 10-14) describes a pattern-matching algorithm for a value pattern.
The body of this match clause compares the target and the value inside the value pattern.
The \lstinline{match-first} expression that recursively calls the \lstinline{Multiset} matcher is used for this comparison.

The fourth match clause (line 15) describes a pattern-matching algorithm for a pattern variable and wildcard.
This match clause creates the next matching atom by just changing the matcher from ``\lstinline{(Multiset M)}'' to \lstinline{Something}.

\subsection{Compiling \lstinline{match-all} to an Application of \lstinline{map}}\label{method-map}

The \lstinline{match-all} expression is transformed into an application of \lstinline{map} whose first argument is a function whose argument is a return value of \lstinline{extract-pattern-variables}, and second argument is the result of \lstinline{gen-match-results}.
The \lstinline{extract-pattern-variables} function takes a pattern and returns a list of pattern variables that appear in the pattern.
The order of the pattern variables corresponds with the order they appeared in the pattern.
For example, ``\lstinline{(extract-pattern-variables '(cons x xs)}'' returns ``\lstinline{(x xs)}''.
The \lstinline{gen-match-results} function takes a pattern, a target, and a matcher, and returns a list of pattern-matching results.
These pattern-matching results consist of values that are going to be bound to the pattern variables returned by \lstinline{extract-pattern-variables}.
The order of the values in the \lstinline{gen-match-results} must correspond with the order of pattern variables returned by \lstinline{extract-pattern-variables}.
For example, ``\lstinline{(gen-match-results '(cons x xs) (Multiset Something) '(1 2 3))}'' returns ``\lstinline{((1 (2 3)) (2 (1 3)) (3 (1 2)))}''.

\begin{lstlisting}[language=egison]
(match-all t M [p e])
;=> `(map (lambda ,(extract-pattern-variables p) ,e) (gen-match-results ,p ,M ,t))
\end{lstlisting}

\noindent The above transformation is done by the following macro.

\begin{lstlisting}[language=egison]
(define-macro (match-all t M . clauses)
  (if (eq? clauses '())
      '()
      (let* {[clause (car clauses)]
             [p (rewrite-pattern (list 'quasiquote (car clause)))]
             [es (cdr clause)]}
        `(append (map (lambda (ret) (apply (lambda ,(extract-pattern-variables p) . ,es) ret))
                      (gen-match-results ,p ,M ,t))
                 (match-all ,t ,M . ,(cdr clauses))))))
\end{lstlisting}

\noindent
\lstinline{define-macro} in the above program is a traditional non-hygienic macro similar to \lstinline{defmacro} of Common Lisp~\cite{gaucheTM}.
The \lstinline{match-all} macro requires to call the complex Scheme procedures such as \lstinline{extract-pattern-variables} for expanding the macro.
Therefore, defining the macro just by using \lstinline{syntax-rules}, which is supported in R7RS~\cite{shinn2013revised} is difficult.
By using \lstinline{syntax-case} and \lstinline{syntax->datum} in R6RS~\cite{sperber2009revised}, we can define the above macro relatively easily.

The program below defines \lstinline{gen-match-results}.
\lstinline{gen-match-results} is a function that calls the \lstinline{processMStates} function introduced in Sect.~\ref{algorithm}.
An initial matching state ``\lstinline|(MState {[p M t]} {})|'' is created from ``\lstinline{p}'', ``\lstinline{M}'', and ``\lstinline{t}''.

\begin{lstlisting}[language=egison]
(define gen-match-results
  (lambda (p M t)
    (processMStates `{(MState {[,p ,M ,t]} {})})))
\end{lstlisting}

\medskip

Generally, pattern matching for algebraic data types is compiled to an expression that uses only more primitive conditional branches such as the \texttt{if} expressions~\cite{peyton1987implementation}.
As a result, the compiled program is as efficient as the program that was written manually without using pattern matching.  
However, the pattern-matching procedure of Egison is so complicated that it is impossible to compile pattern-matching directly to an expression only with more primitive conditional branches.
This section showed that we can compile the large pattern-matching procedure of Egison by separating the extraction of pattern variables and the calculation of pattern-matching results.
These two procedures are implemented in Scheme, and a pattern-matching expression is compiled to a program that calls these two procedures.
This method allows us to compile very complicated pattern matching procedures but has overhead.
Sect.~\ref{benchmark-result} analyzes this overhead.

\subsection{Compiling Value Patterns to \lstinline{lambda}s}\label{method-val-pat}

A value pattern is transformed into a \lstinline{lambda} expression whose arguments are pattern variables that appear in the left side of the value pattern.
For example, ``\lstinline{(cons x (cons y (cons ,x (cons z _))))}'' is transformed into ``\lstinline{(cons x (cons y (cons (val ,(lambda (x y) x)) (cons z _))))}''.
The \lstinline{rewrite-pattern} function called inside the macro does this transformation.

This transformed value pattern is handled in \lstinline{processMState}.
The following program shows a match clause in \lstinline{processMState} for handling a value pattern, which is omitted in the explanation of \lstinline{processMState} in Sect.~\ref{algorithm}.

\begin{lstlisting}[language=egison]
(define processMState
  (lambda (mState)
    (match mState
      [('MState {[('val f) M t] . mStack} ret)
       (let {[next-matomss (M `(val ,(apply f ret)) t)]}
         (map (lambda (next-matoms) `(MState ,(append next-matoms mStack) ,ret)) next-matomss))]
      ...)))
\end{lstlisting}

\noindent
When the pattern of the top matching atom is a value pattern, it applies the intermediate pattern-matching result to the function in the value pattern and passes it to the matcher.
This pattern is handled by the third match clause of the \lstinline{Multiset} matcher in Sect.~\ref{method-matcher}, for example.

\section{Performance}\label{performance}

The section discusses the performance of the proposed Scheme macros.
Sect.~\ref{tip} presents an idea for improving the performance of Egison pattern matching that was implemented for the first time in the proposed Scheme macros.
Sect.~\ref{benchmark-result} shows the benchmark results of Egison and the proposed Scheme macros and analyzes these benchmark results.

\subsection{Tips for Improving Performance}\label{tip}

There are cases that the performance of pattern matching improves by adding a new match clause to a matcher.
This subsection introduces such examples.

First, we can improve the performance of the \lstinline{Multiset} implementation in Sect.~\ref{method-matcher} using this method.
The \lstinline{Multiset} matcher definition below has a new match clause in the 6th line.
This match clause describes a pattern-matching algorithm for the \lstinline{cons} pattern whose second argument is a wildcard.
In this case, we do not need to calculate the rest of the collection.
The cost of calculating the rest of the collection is so high that the performance dramatically improves by adding this new match clause.

\begin{lstlisting}[language=egison]
(define Multiset
  (lambda (M)
    (lambda (p t)
      (match p
        ...
        [('cons px '_) (map (lambda (x) `{[,px ,M ,x]}) t)]
        [('cons px py)
         (map (lambda (xy) `{[,px ,M ,(car xy)] [,py ,(Multiset M) ,(cadr xy)]})
              (match-all t (List M)
                [(join hs (cons x ts)) `(,x ,(append hs ts))]))]
        ...))))
\end{lstlisting}

Here is another example from the \lstinline{List} matcher.
The cost of pattern matching for the \lstinline{join} pattern is high.
However, the cost for enumerating only the tail parts is low; we can write a tail-recursive function for enumerating the tail parts.
When we use join patterns, the first argument of \lstinline{join} is often a wildcard.
Therefore, by adding a match clause for a \lstinline{join} pattern whose first argument is a wildcard as follows, we can improve the efficiency of pattern matching for a list.

\begin{lstlisting}[language=egison]
(define List
  (lambda (M)
    (lambda (p t)
      (match p
        [('nil) (if (eq? t '{}) '{{}} '{})]
        [('cons px py)
         (match t
                (() '{})
                ((x . xs)
                 `{{[,px ,M ,x] [,py ,(List M) ,xs]}}))]
        [('join '_ py)
         (map (lambda (y) `{[,py ,(List M) ,y]})
               (tails t))]
        [('join px py)
         (map (lambda (xy) `{[,px ,(List M) ,(car xy)] [,py ,(List M) ,(cadr xy)]})
              (unjoin t))]
        [('val x) (if (eq? x t) '{{}} '{})]
        [pvar `{{[,pvar Something ,t]}}]))))
\end{lstlisting}

\subsection{Benchmark Result}\label{benchmark-result}

I compared the performance of a nested \texttt{cons} pattern for a multiset.
For benchmarking Egison and the proposed Scheme macros, I used the expressions below.
\texttt{between} is a function of Egison that returns a list that contains sequential integers between the first and second arguments.

\begin{lstlisting}[language=egison]
(match-all (between 1 n) (multiset something) [<cons $x <cons $y _>> [x y]])
\end{lstlisting}

\noindent
Similarly, \texttt{iota} is a function in Scheme that returns a list of sequential integers that starts from the integer of the second argument.

\begin{lstlisting}[language=egison]
(match-all (iota n 1) (Multiset Something) [(cons x (cons y _)) `(,x ,y)])
\end{lstlisting}

\noindent
I also made a comparison with the functional description of the same program in Scheme.

\begin{lstlisting}[language=egison]
(define comb2 (lambda (xs) (comb2-helper xs '{})))

(define comb2-helper
  (lambda (xs hs)
    (if (eq? xs '{})
        '{}
        (append (append (map (lambda (y) `(,(car xs) ,y)) hs)
                        (map (lambda (y) `(,(car xs) ,y)) (cdr xs)))
                (comb2-helper (cdr xs) (append hs `(,(car xs))))))))

(comb2 (iota n 1))
\end{lstlisting}

\noindent
I used Egison version 3.8.1 and Gauche version 0.9.7 on 2.3 GHz Intel Core i5 processor for this benchmark.

\begin{table}[htbp]
\begin{tabular}{|l||l|l|l|l|l|l|} \hline
\texttt{comb2} & n=50 & n = 100 & n = 200 & n=400 & n=800 & n=1600 \\ \hhline{|=#=|=|=|=|=|=|}
Egison with \texttt{multiset} in Sect.~\ref{method-matcher} & 1.189s & 4.470s & 21.441s & 112.67s & 697.66s & n/a \\ \hline
Egison with \texttt{multiset} in Sect.~\ref{tip} & 0.438s & 1.312s & 4.751s & 22.612s & 112.89s & 714.43s \\ \hline
Scheme with \texttt{Multiset} in Sect.~\ref{method-matcher} & 0.124s & 0.436s & 2.413s & 17.900s & 117.41s & n/a \\ \hline
Scheme with \texttt{Multiset} in Sect.~\ref{tip} & 0.074s & 0.099s & 0.259s & 0.752s & 3.159s & 12.641s \\ \hline
Scheme (functional style) & 0.026s & 0.042s & 0.107s & 0.373s & 1.618s & 7.949s \\ \hline
\end{tabular}
\caption{Benchmark results}
\label{table:benchmark}
\end{table}


Table~\ref{table:benchmark} shows the benchmark results.
I ran all the benchmarks five times and took a median for each of them.
For all the benchmarks, Scheme is faster than Egison.
The reason for these performance differences is that Gauche, a scheme implementation I used for the benchmark, has a compiler while Egison has only a simple interpreter implementation.

Adding a matcher clause for a constructor pattern with wildcards is effective for both Egison and Scheme.
This trick improves the performance of Egison and Scheme around $6$ times and $40$ times, respectively for $n=800$.
The performance improvement is larger for Scheme than Egison because Scheme is a strict language.
Scheme always evaluates ``\lstinline{(append hs ts)}'' in the 9th line of the \lstinline{Multiset} definition in Sect.~\ref{method-matcher} whereas Egison evaluates the corresponding expression only when it is necessary.

We can also observe that pattern-matching-oriented programming style is only two times slower than the functional programming style in this case, though the program in pattern-matching-oriented programming style much simpler.

\section{Future Work}\label{future}

The implementation of more expressive pattern constructs for non-free data types remains as future work.
For example, several advanced pattern constructs in Egison such as \emph{loop patterns}~\cite{egi2018loop}, \emph{sequential patterns}, and \emph{pattern functions}~\cite{egisonPat} are still not implemented in the proposed Scheme macros.
Loop patterns allow us to describe repetitions in a pattern like the Kleene star operator of the regular expressions.
Loop patterns are more expressive than the other repeated pattern such as the Kleene star operator in the sense that loop patterns allow users to change the pattern repeated depending on the current repeat count~\cite{egi2018loop}.
Sequential patterns are a generalization of later patterns and allow users to control the order of the pattern-matching process.
Pattern functions allow users to modularize patterns with a lexical scope by composing patterns.
These pattern constructs widen the range of patterns that we can describe concisely.
Currently, I am preparing a paper to show the usefulness of these patterns.

The integration of Egison pattern matching with a programming language with a static type system also remains as future work.
Though we have an experimental Template Haskell implementation~\cite{egisonHaskell}, this Haskell implementation currently has a limitation that it does not allow users to define new pattern constructors.
For example, the \lstinline{nil}, \lstinline{cons}, and \lstinline{join} pattern constructors are built-in in the current implementation.
This limitation comes from the limitation of Haskell that does not allow overlapping data families~\cite{ghcMan}.
In the current Template Haskell implementation of Egison, patterns are defined in the library as data of ``\lstinline{Pattern a}'', which is a data type for patterns against data whose type is ``\lstinline{a}''.
Users cannot add new data constructors in ``\lstinline{Pattern a}'' in their program for adding new pattern constructors because of overlap is not permitted.
Currently, we are working to find a method for avoiding this problem.

\section{Conclusion}\label{conclusion}

This paper proposed a method for compiling Egison pattern-matching expressions that can describe very expressive pattern matching for non-free data types but has a complicated pattern-matching procedure.
The proposed method made an implementation of the pattern-matching system of Egison very short; the macros and the pattern-matching procedure is implemented in less than 150 lines of Scheme code.
This implementation is also a good proof of the extensibility of Scheme and Lisp.

This pattern-matching library will help the research of pattern matching by providing an easy method for developing new pattern-matching extensions.
This pattern-matching library will also help the education of algorithms by making its implementation simpler as the implementation of a SAT solver presented in this paper.
I hope this work leads to the further development and propagation of advanced pattern matching.

\begin{acks}
  I thank Yuichi Nishiwaki for constructive discussion while implementing the proposed Scheme macros and writing this paper.
  I thank Tatsuya Yamashita for the first implementation of a SAT solver in Egison.
  I thank Yuito Murase for implementing the proposed macros in Common Lisp.
  I thank Mayuko Kori for implementing the proposed macros using Template Haskell.
  Finally, I greatly thank my shepherd William Byrd and the anonymous reviewers of the 20th Scheme and Functional Programming Workshop for their reviews that dramatically improved this paper.
\end{acks}

\bibliographystyle{ACM-Reference-Format}
\bibliography{main}

\newpage

\appendix

\section{Pattern-Matching-Oriented Programming}\label{apps}

The pattern-matching facility of Egison is so expressive that a new programming style called \emph{pattern-matching-oriented programming style} arises.
This programming style makes descriptions of mathematical algorithms concise by confining explicit recursions for traversing data inside intuitive patterns.
This section shows several sample programs in this programming style.
I am currently preparing another paper for discussing pattern-matching-oriented programming techniques.
We can execute programs shown in this section by loading a program in \cite{egisonScheme} on Gauche~\cite{gaucheWeb}.

\subsection{Basic List Processing Functions in Pattern-Matching-Oriented Style}\label{pmo-list}

Pattern matching for non-free data types enables more intuitive definitions of even the basic list processing functions such as \lstinline{map}, \lstinline{concat}, and \lstinline{unique} by confining recursion inside a pattern.

The \lstinline{map} function is defined using pattern matching as follows.
The \lstinline{(join _ (cons x _))} pattern matches each element of a target list.
We call this pattern \emph{join-cons pattern} because it often appears in list programming.
Combining a join-cons pattern with \lstinline{match-all}, we can simply implement the \lstinline{map} function.

\begin{lstlisting}[language=egison]
(define pm-map
  (lambda (f xs)
    (match-all xs (List Something)
      [(join _ (cons x _)) (f x)])))

(pm-map (lambda (x) (+ x 10)) `(1 2 3 4))
; (11 12 13 14)
\end{lstlisting}

\noindent By doubly nesting the above join-cons pattern, we can define the \lstinline{concat} function.
Note that, we can create a matcher for such as a list of lists and a multiset of multisets by composing matchers.

\begin{lstlisting}[language=egison]
(define pm-concat
  (lambda (xss)
    (match-all xss (List (List Something))
      [(join _ (cons (join _ (cons x _)) _)) x])))

(pm-concat `((1 2) (3) (4 5)))
; (1 2 3 4 5)
\end{lstlisting}

By combining a not-pattern with a doubly-nested join-cons pattern, we can also define a \lstinline{unique} function.
The pattern below extracts only the last appearance of each element; a not-pattern is used to describe that there is no more \lstinline{x} after an occurrence of \lstinline{x}.
\texttt{Eq} is a matcher that can handle pattern matching for a value pattern.
\texttt{eq?} is used for checking equality.

\begin{lstlisting}[language=egison]
(define pm-unique-simple
  (lambda (xs)
    (match-all xs (List Eq)
      [(join _ (cons x (not (join _ (cons ,x _))))) x])))

(pm-unique-simple `(1 2 3 2 4))
; (1 3 2 4)
\end{lstlisting}

\noindent We can define \lstinline{unique} whose results consist of the first appearance of each element by using a later pattern.
A later pattern is used to describe that there is no appearance of \texttt{x} in the target list for the first argument of the \texttt{join} pattern.

{\footnotesize
\begin{lstlisting}[language=egison]
(define pm-unique
  (lambda (xs)
    (match-all xs (List Eq)
      [(join (later (not (join _ (cons ,x _)))) (cons x _)) x])))

(pm-unique `(1 2 3 2 4))
; (1 2 3 4)
\end{lstlisting}
}

\subsection{Implementation of a SAT Solver}

The program below describes the Davis-Putnam algorithm.
We can see a full implementation of this SAT solver in \texttt{dp.scm} in the GitHub repository of the proposed Scheme macros~\cite{egisonScheme}.
Pattern matching for multisets dramatically improves the readability of the description of this algorithm.
We can compare this Scheme program with the OCaml implementation of the same algorithm in~\cite{harrison2009handbook}.

The \lstinline{sat} function takes two argument \lstinline{vars}, a list of propositional variable, and \lstinline{cnf}, a propositional logic formula in conjunctive normal form.
The tuple consists of \lstinline{vars} and \lstinline{cnf} is pattern-matched as a tuple of a multiset of integers and a multiset of multisets of integers in the definition of \lstinline{sat}.
In this program, a propositional variable is represented as a positive integer, and a literal is represented as an integer.
For example, $-1$ represents negation of the propositional variable $1$.

\begin{lstlisting}[language=egison]
(define sat
  (lambda [vars cnf]
    (match-first `[,vars ,cnf] `[,(Multiset Integer) ,(Multiset (Multiset Integer))]
      ; satisfiable
      ['[_ ()] #t]
      ; unsatisfiable
      ['[_ (cons () _)] #f]
      ; 1-literal rule
      ['[_ (cons (cons l ()) _)]
       (sat (delete (abs l) vars) (assign-true l cnf))]
      ; pure literal rule (positive)
      ['[(cons v vs) (not (cons (cons ,(neg v) _) _))]
       (sat vs (assign-true v cnf))]
      ; pure literal rule (negative)
      ['[(cons v vs) (not (cons (cons ,v _) _))]
       (sat vs (assign-true (neg v) cnf))]
      ; otherwise
      ['[(cons v vs) _]
       (sat vs (append (resolve-on v cnf)
               (delete-clauses-with v (delete-clauses-with (neg v) cnf))))])))
\end{lstlisting}

\noindent
The first match clause (line 5) matches if \lstinline{cnf} is empty.
This match clause returns \lstinline{#t} because \lstinline{cnf} is satisfiable in this case.
The second match clause (line 7) matches if \lstinline{cnf} contains an empty clause.
This match clause returns \lstinline{#f} because \lstinline{cnf} is unsatisfiable in this case.
The third match clause (lines 9 and 10) matches if \lstinline{cnf} contains a clause that consists of a single literal \lstinline{l}.
In this match clause, we assign \lstinline{l} true at once.
The fourth match clause (lines 12 and 13) matches when there is a propositional variable \lstinline{v} that appears only positively in \lstinline{cnf}.
In this match clause, we assign \lstinline{v} true at once.
The fifth match clause (lines 15 and 16) matches when there is a propositional variable \lstinline{v} that appears only negatively in \lstinline{cnf}.
In this match clause, we assign \lstinline{v} false at once.
The final match clause (lines 18-20) matches when pattern matching for all the above match clauses fails.
This match clause applies the resolution principle.

\section{Pattern Matching with Infinitely Many Results}\label{pm-inf}

\lstinline{match-all} provided by \texttt{stream-egison.scm} supports pattern matching with infinitely many results.
A library for streams described in SRFI 41~\cite{bewig2007scheme} is used for handling lazy lists.
The following \lstinline{match-all} expressions extract an infinite list of twin primes and prime triplets from a stream of prime numbers.

\begin{lstlisting}[language=egison]
(load "./stream-egison.scm")

(define stream-primes (stream-filter bpsw-prime? (stream-iota -1 1)))

(stream->list
 (stream-take
  (match-all stream-primes (List Integer)
             [(join _ (cons p (cons ,(+ p 2) _)))
              `(,p ,(+ p 2))])
  10))
; ((3 5) (5 7) (11 13) (17 19) (29 31) (41 43) (59 61) (71 73) (101 103) (107 109))

(stream->list
 (stream-take
  (match-all stream-primes (List Integer)
             [(join _ (cons p (cons (and (or ,(+ p 2) ,(+ p 4)) m) (cons ,(+ p 6) _))))
              `(,p ,m ,(+ p 6))])
  8))
; ((5 7 11) (7 11 13) (11 13 17) (13 17 19) (17 19 23) (37 41 43) (41 43 47) (67 71 73))
\end{lstlisting}

\noindent
\lstinline{bpsw-prime?} is a predicate that checks whether the argument is a prime number or not.
This predicate is provided by the \texttt{math.prime} Gauche library~\cite{gauchePN}.
\lstinline{stream-iota}, \lstinline{stream->list}, and \lstinline{stream-take} are provided by the \texttt{util.stream} Gauche library~\cite{gaucheST}.
``\lstinline{(stream-iota -1 1)}'' returns a stream that contains all the positive integers.
\lstinline{stream-take} returns a stream that contains the first elements of the first argument stream.
The number of elements taken from the given stream is specified by the second argument.
\lstinline{stream->list} is a function that converts the given stream to a list.

\begin{lstlisting}[language=egison]
(stream->list (stream-take (stream-iota -1 1) 10))
; (1 2 3 4 5 6 7 8 9 10)
\end{lstlisting}

In the languages whose default evaluation strategy is non-strict as Haskell and Egison, we do not need to distinguish these two \lstinline{match-all} implementations.

\end{document}